\documentclass{hep99}
\usepackage{cite}
\def\be{\begin{equation}}
\def\ee{\end{equation}}
\def\bea{\begin{eqnarray}}
\def\eea{\end{eqnarray}}
\def\pom{I\!\!P}
\begin{document}
\frenchspacing

\title{Review of different approaches to understanding the phenomenon of 
diffraction}

\author{A. Hebecker}

\address{Institut f\"ur Theoretische Physik der Universit\"at Heidelberg\\
Philosophenweg 16, D-69120 Heidelberg, Germany}

\abstract{This brief review of hard diffraction is focussed on the 
theory of the diffractive structure function $F_2^D$. Some aspects of 
diffractive vector meson production and of diffractive processes in 
hadron-hadron collisions are also discussed.} 

\maketitle

\section{Introduction}
With the advent of HERA, deep inelastic scattering (DIS) at very small $x$ 
became experimentally viable. One of the most interesting phenomena 
observed in this kinematical domain are large rapidity gap events. In these 
events, a diffractive final state with mass $M$ is well separated in 
rapidity from the elastically scattered proton or its low-mass excitation. 
At small values of the Bjorken variable $x\simeq Q^2/W^2$ (where $Q^2=-q^2$ 
is the virtuality of the exchanged photon and $W^2$ is the invariant mass 
square of the photon-proton collision) these events form a leading twist 
contribution to DIS, which is described by the diffractive structure 
function $F_2^D$~(for recent results see~\cite{f2d}). 

Hard diffraction was previously observed in hadron-hadron 
collisions~\cite{ua8}. However, diffractive DIS has the advantage of being 
simpler since only one initial state hadron is involved. Thus, HERA has 
triggered considerable renewed interest in theoretical approaches to 
diffraction. 

This paper is organized as follows. In Sect.~\ref{ajm} diffractive DIS
is discussed in the rest frame of the target proton. Section~\ref{pp} is 
devoted to the Breit frame, where the proton is fast and a parton model 
description is appropriate. The relation between these two approaches is 
analyzed in Sect.~\ref{vs}. Aspects of diffractive vector meson production 
in DIS and of diffractive processes in hadron-hadron collisions are 
discussed in Sects.~\ref{emp} and \ref{hh} respectively. Some open problems 
are mentioned in Sect.~\ref{op}.

\section{Aligned jet model and its modern versions}\label{ajm}
The aligned jet model~\cite{bk} is based on a qualitative picture of DIS 
in the target rest frame, where the incoming virtual photon can be 
described as a superposition of partonic states. The large virtuality 
$Q^2$ sets the scale, so that states with low-$p_\perp$ partons, i.e., 
aligned configurations, are suppressed in the photon wave function. 
However, in contrast to high-$p_\perp$ configurations, these aligned 
states have a large interaction cross section with the proton. Therefore, 
their contribution to DIS is expected to be of leading twist. Since the 
above low-$p_\perp$ configurations represent transversely extended, 
hadron-like objects, which have a large elastic cross section with the 
proton, part of this leading twist contribution is diffractive. 

The above intuitive picture was implemented in the framework of perturbative 
QCD in~\cite{wf}, where the colour singlet exchange between the proton and 
the $q\bar{q}$ fluctuation of the $\gamma^*$ is realized by two gluons. A 
further essential step is the inclusion of higher Fock states in the photon 
wave function. In the framework of two gluon exchange, corresponding 
calculations for the $q\bar{q}g$ state were performed in~\cite{nz}. 
The main shortcoming of the two-gluon approach is the lacking justification 
of perturbation theory. As should be clear from the qualitative discussion 
of the aligned jet model, the diffractive kinematics is such that the $t$ 
channel colour singlet exchange does not feel the hard scale of the 
initial photon. Thus, more than two gluons can be exchanged without 
suppression by powers of $\alpha_s$. 

This problem can be systematically addressed in the semiclassical 
approach~\cite{bh}, where the interaction with the target is modelled as 
the scattering off a superposition of soft colour fields. In the 
high-energy limit, the eikonal approximation can be used. Diffraction 
occurs if both the target and the partonic fluctuation of the photon remain 
in a colour singlet state. Thus, both the diffractive and inclusive DIS 
cross section can be calculated if a model for the wave functional of the 
proton is provided.

\section{From partonic pomeron to diffractive parton distribution 
functions}\label{pp}
It is tempting to interpret the quasi-elastic high-energy scattering of 
photon fluctuation and proton in terms of pomeron exchange, thus introducing 
a soft energy dependence. A more direct way of applying the concept of the 
soft pomeron to hard diffraction was suggested in~\cite{is}. Essentially, 
one assumes that the pomeron can, like a real hadron, be characterized by 
a parton distribution. This distribution is assumed to factorize from the 
pomeron trajectory and the pomeron-proton-proton vertex, which are both 
obtained from the analysis of purely soft hadronic reactions. The problem 
with this approach is the lacking justification of the pomeron idea and 
the factorization assumption in QCD. 

The concept of fracture functions~\cite{vt} or, more specifically, the 
diffractive parton distributions of~\cite{bs} provide a framework for the 
study of diffractive DIS that is firmly rooted in perturbative QCD. 
In short, diffractive parton distributions are conditional probabilities. 
A diffractive parton distribution $df^D_i(y,\xi,t)/d\xi\,dt$ describes the 
probability of finding, in a fast moving proton, a parton $i$ with momentum 
fraction $y$, under the additional requirement that the proton 
remains intact while being scattered with invariant momentum transfer $t$ 
and losing a small fraction $\xi=x_{\pom}$ of its longitudinal momentum. 
Thus, the corresponding $\gamma^*p$ cross section can be written 
as~\cite{bs1}\\[.0cm]
\be
\hspace*{-4cm}\frac{d\sigma(x,Q^2,\xi,t)^{\gamma^*p\to p'X}}{d\xi\,dt}
\label{sx}
\ee
\[
\hspace{1cm}=\sum_i\int_x^\xi 
dy\,\hat{\sigma}(x,Q^2,y)^{\gamma^*i}\left(\frac{df^D_i(y,\xi,t)}{d\xi\,dt}
\right)\, ,
\]
where $\hat{\sigma}(x,Q^2,y)^{\gamma^*i}$ is the total cross section for 
the scattering of a virtual photon characterized by $x$ and $Q^2$ and a 
parton of type $i$ carrying a fraction $y$ of the proton momentum. The 
above factorization formula holds in the limit $Q^2\to\infty$ with $x$, 
$\xi$ and $t$ fixed. Factorization proofs were given in~\cite{gtv} in the 
framework of a simple scalar model and in~\cite{cfa} in full QCD. 

As in inclusive DIS, there are infrared divergences in the partonic cross 
sections and ultraviolet divergences in the parton distributions. Thus, a 
dependence on the factorization scale $\mu$ appears both in the parton 
distributions and in the partonic cross sections. The claim that 
Eq.~(\ref{sx}) holds to all orders implies that these $\mu$ dependences 
cancel, as is well known in the case of conventional parton distributions. 
Therefore, the diffractive distributions obey the usual DGLAP evolution 
equations.

\section{Target rest frame vs. Breit frame}\label{vs} 
It is instructive to compare the two different approaches to diffractive 
DIS presented in Sections~\ref{ajm} and~\ref{pp}, which correspond to the 
target rest frame and the Breit frame respectively. The relation between 
these different points of view was discussed in~\cite{nz,afs}. An explicit 
calculation demonstrating how diffractive parton distributions arise from 
a target rest frame perspective was given in~\cite{h}. 

In the rest frame of the proton, the incoming $\gamma^*$ fluctuates into 
a $q\bar{q}$ pair before it reaches the target. The leading twist 
diffractive contribution comes from small-$p_\perp$, aligned 
configurations. These configurations are asymmetric, i.e., one of the 
two quarks carries most of photon momentum, while the other quark is 
relatively soft. Kinematically, the energetic quark does not feel the 
soft colour fields of the proton. Boosting this physical picture to a 
frame where the proton is fast, e.g., the Breit frame, the soft quark 
changes its direction: now the fast colour field of the proton creates a 
colour singlet $q\bar{q}$ pair; one of the quarks scatters off the 
$\gamma^*$ in a familiar partonic process ($\gamma^*q\to q$); the produced 
quark is treated as a free final state particle. This quark corresponds to 
the energetic quark from the $q\bar{q}$ fluctuation of the photon in the 
target rest frame. 

Analogously, the process based on a $q\bar{q}g$ fluctuation of the 
$\gamma^*$ can be reinterpreted in the Breit frame as boson-gluon fusion 
with the gluon coming from the diffractive gluon distribution. 

The above picture shows that models for the proton color field can be 
translated into diffractive parton distributions at some low starting 
scale, and a standard DGLAP analysis can be used to compare to data. In 
ref.~\cite{hks}, a small colour dipole was used as a model for the proton 
and diffractive parton distributions were calculated using operator 
definitions. In ref.~\cite{bgh}, the analysis was based on formulae for 
parton distributions from~\cite{h} and a large hadron model was used. 
Both analyses describe the data well. In spite of the different models 
for the target, common qualitative features emerge: in both cases the 
gluon distribution is dominant and falls off in the limit $y/\xi\to 1$.

\section{Energy growth; higher twist; final states}
The energy growth of diffraction, i.e., the behaviour in the limit 
$\xi\to 0$, is one of the less well understood aspects of the process. 
Phenomenologically, the small-$\xi$ behaviour of $F_2^D$ is very similar 
to the small-$x$ behaviour of $F_2$~\cite{b}. Attributing the energy 
growth to the soft colour field dynamics of the proton, this relation 
can be easily understood in the semiclassical approach~\cite{bh}. However, 
no explicit derivation of the $\xi$ dependence has been given. 

Motivated by the idea of saturation, an energy or $x$ dependence in a 
Glauber-type expression for the dipole cross section $\sigma(\rho)$ is 
introduced in~\cite{gbw}. A similar energy dependence of diffractive and 
inclusive DIS is found. 

In perturbation theory, the growth of the diffractive cross section for 
$\xi\to 0$ is ascribed to BFKL dynamics. However, the applicability of 
perturbative methods is questionable and naive perturbative estimates tend 
to give a small-$\xi$ behaviour that is too steep. Nevertheless, with a 
certain number of parameters a good description of the data can be 
achieved~\cite{mpr}. 

Higher twist effects in $F_2^D$ are particularly important in the 
region where $\beta=x/\xi\to 1$. This corresponds to the well-known 
higher-twist problem in inclusive DIS at large $x$ (for phenomenological 
implications of higher twist contributions see, e.g.,~\cite{ht}). 

It has been proposed to use final states with charm or high-$p_\perp$ jets 
to keep the $t$ channel colour singlet exchange hard and therefore 
calculable in the two-gluon approximation (see, e.g.,~\cite{ccpt}). 
However, this approach has the problem that high-$p_\perp$ or charmed 
final states can also arise from $q\bar{q}g$ fluctuations of the 
$\gamma^*$, which are non-perturbative as far as the $t$ channel exchange 
is concerned.

\section{Elastic vector meson production}\label{emp}
Diffractive processes where the $t$ channel colour singlet exchange is 
governed by a hard scale include the electroproduction of heavy vector 
mesons~\cite{rys}, electroproduction of light vector mesons in the case of 
longitudinal polarization~\cite{bro} or at large $t$~\cite{fr}, and virtual 
Compton scattering~\cite{mea,ji}. In the leading logarithmic 
approximation, the relevant two-gluon form factor of the proton can be 
related to the inclusive gluon distribution~\cite{rys}. Accordingly, a very 
steep energy dependence of the cross section, which is now proportional to 
the square of the gluon distribution, is expected.

To go beyond leading logarithmic accuracy, the non-zero momentum transferred 
to the proton has to be taken into account. This requires the use of 
`non-forward' or `off-diagonal' parton distributions (see~\cite{mea} and 
refs. therein), which were discussed in~\cite{ji} within the present 
context. Although their scale dependence is predicted by well-known 
evolution equations, only limited information about the relevant input 
distributions is available (see, however,~\cite{fnf} for possibilities of 
predicting the non-forward from the forward distribution functions). 

The perturbative calculations of meson electroproduction discussed above 
were put on a firmer theoretical basis by the factorization proof of 
\cite{cfs1}. It was also shown that in the case of transverse polarization 
the cross section is suppressed by a power of $Q^2$ in the high-$Q^2$ 
limit. However, present data do not support this expectation. 

As an alternative to the description of the final state meson in terms 
of a light-cone wave function, it was suggested to approach both 
transversely and longitudinally polarized diffractive $\rho$ production on 
the basis of open $q\bar{q}$ production combined with the idea of 
parton-hadron duality~\cite{mrt}. A non-perturbative treatment of the $t$ 
channel exchange is used in~\cite{do}. 

An interesting new field, closely related to exclusive vector meson 
production, are the semi-exclusive processes discussed in~\cite{sep}.

\section{Diffraction in hadron-hadron collisions}\label{hh}
The two main types of hard diffraction in hadron-hadron scattering are 
high-$p_\perp$ jet production where one or both incoming hadrons remain 
intact and high-$p_\perp$ jets with a rapidity gap between the jets. 

The first of these two processes is the original process of hard 
diffraction~\cite{ua8} for which the concept of the partonic pomeron was 
designed~\cite{is}. However, the underlying factorization assumption is 
not expected to be precise in QCD (cf. the model calculation of~\cite{bs}). 
Indeed, the attempt to to apply diffractive parton distributions measured 
at HERA to hard diffraction at the Tevatron shows very substantial 
factorization breaking~\cite{alv}. 

Intuitively, the breakdown of diffractive factorization in hadron-hadron 
collisions is easily understood since the remnants of the initial state 
hadrons can interact with the final state particles. This is in contrast to 
the DIS case, where no $\gamma^*$ remnant exists. Concepts like `gap 
survival probability' and `pomeron flux renormalization' have been invoked 
for the description of the effect~\cite{gsp}, but no accepted theoretical 
approach has emerged so far. 

Processes with two high-$p_\perp$ jets and a large gap without hadronic 
activity in between have been investigated as a possible probe for BFKL 
dynamics~\cite{gbj}. The main idea is that processes of this type proceed 
dominantly with colour singlet exchange between the high-$p_\perp$ partons. 
One faces again the gap survival problem. Recently, it has been 
proposed to avoid this problem by defining the gap in terms of transverse 
energy flow rather than in terms of multiplicity~\cite{os}. 

Note also that in a recent development of the soft-colour Monte Carlo 
method the colour exchange is described on the level of strings rather 
than on the partonic level~\cite{rath}, and a combined description 
of hard diffraction in DIS and in hadron-hadron collisions is 
achieved~\cite{ing}.

\section{Open problems}\label{op}
One of the most interesting problems in diffractive DIS is the asymptotic 
energy or $\xi$ dependence of $F_2^D$ and its possible relation to the 
small-$x$ limit of DIS. There is good reason to expect a close connection 
between these two limits, but no definitive answer has yet been provided. 
Furthermore, an analysis of $F_2^D$ including both DGLAP evolution and 
higher twist effects is urgently needed. In vector meson production, the 
apparent disagreement of the QCD expectation of a $1/Q^2$ suppression of 
the transverse cross section with data has to be explained. Finally, a 
quantitative, QCD based understanding of the gap survival problem in 
hadron-hadron collisions has to be developed.

\section*{Acknowledgements}
I would like to thank the organizers of EPS-HEP 99 and, in particular, the 
conveners of the diffractive session, V.~Del~Duca and M.~Erdmann, for their 
invitation to present this review.

\end{document}